\documentclass[preprint]{elsarticle}

\usepackage{lineno,hyperref}

\usepackage[inline]{enumitem}
\usepackage[table,xcdraw]{xcolor}
\usepackage{longtable}
\usepackage{booktabs}
\usepackage{lscape}
\usepackage{amsmath}
\usepackage{algorithm}
\usepackage{algpseudocode}
\usepackage{multirow}
\usepackage{verbatim}
\usepackage{graphicx}
\usepackage{todonotes}
\usepackage[gen]{eurosym}
\usepackage{tabularx}
\usepackage{xspace}
\usepackage{arydshln}
\newcommand{\algrule}[1][.2pt]{\par\vskip.5\baselineskip\hrule height #1\par\vskip.5\baselineskip}
\algnewcommand{\TRUE}{\textbf{True}}
\algnewcommand{\FALSE}{\textbf{False}}

\DeclareMathOperator*{\argmax}{argmax} 
\newcommand{\crbCRS}{\textsc{CRB-CRS}\xspace}

\presetkeys%
    {todonotes}%
    {inline,backgroundcolor=yellow}{}

\makeatletter
\def\ps@pprintTitle{%
 \let\@oddhead\@empty
 \let\@evenhead\@empty
 \def\@oddfoot{\centerline{~}}%
 \let\@evenfoot\@oddfoot}
\makeatother
\begin{document}

\begin{frontmatter}

\title{Towards Retrieval-based Conversational Recommendation}

\author{Ahtsham Manzoor}
\author{Dietmar Jannach}
\address{University of Klagenfurt, Austria\\email: ahtsham.manzoor@aau.at, dietmar.jannach@aau.at}

\begin{abstract}
Conversational recommender systems have attracted immense attention recently. The most recent approaches rely on neural models trained on recorded dialogs between humans, implementing an end-to-end learning process. These systems are commonly designed to generate responses given the user's utterances in natural language. One main challenge is that these generated responses both have to be appropriate for the given dialog context and must be grammatically and semantically correct. An alternative to such generation-based approaches is to retrieve responses from pre-recorded dialog data and to adapt them if needed. Such retrieval-based approaches were successfully explored in the context of general conversational systems, but have received limited attention in recent years for CRS. In this work, we re-assess the potential of such approaches and design and evaluate a novel technique for response retrieval and ranking. A user study (N=90) revealed that the responses by our system were on average of higher quality than those of two recent generation-based systems. We furthermore found that the quality ranking of the two generation-based approaches is not aligned with the results from the literature, which points to open methodological questions. Overall, our research underlines that retrieval-based approaches should be considered an alternative or complement to language generation approaches.\footnote{This work continues and significantly extends our recent research in \cite{manzoorrs2021}.}
\end{abstract}

\begin{keyword}
Recommender Systems, Conversational Recommendation, User Study, Information Retrieval, End-to-end Learning, Dialog Systems
\end{keyword}

\end{frontmatter}


\section{Introduction}
Conversational recommender systems (CRS) are interactive agents that converse with humans in natural language with the goal of
helping them finding relevant content and making decisions.
In recent years, we observed that such systems obtained increased attention due to the spread of voice-controlled devices and advances in natural language processing (NLP) and machine learning in general, e.g., see \cite{jannach2021crscsur, gao2021advances} for recent surveys on the topic.

Many current works in the area of CRS adapt an end-to-end learning process,  e.g., \cite{ren2020crsal,zhou2020towards,zhou2020improving}. One promise of such learning-based approaches is that they avoid the potential knowledge engineering bottleneck of traditional constraint-based or critiquing-based systems like \cite{chen2007preference,recsyshandbook2015constraints,Jannach:2004:ASK:3000001.3000153}
by relying on machine learning models. These models are usually trained on large corpora of recorded recommendation dialogs between humans. The \emph{ReDial} datatset is one of the prominent examples of such a dialog corpus, which was used for CRS development in the movie domain, e.g., in \cite{zhou2020improving,li2018towards,chen-etal-2019-towards}. From a technical perspective, these recent works on CRS are based on a \emph{language generation} approach, where the system makes use of the learned model to generate responses in natural language given the previous user utterances in a dialog. In recent years, different deep neural network architectures were explored for this language generation task with the aim to continuously improve the quality of the generated responses. Such architectures for example include recurrent neural networks (RNN), sequence-to-sequence transformer and encoder-decoder models \cite{li2018towards,chen-etal-2019-towards,DBLP:journals/corr/JoshiMF17,kang2019recommendation, hayati2020inspired}. Furthermore, these networks are sometimes coupled with additional components and external knowledge sources for the recommendation task, e.g., in \cite{zhou2020improving,chen-etal-2019-towards}.

One main challenge of every CRS that supports natural language interaction is to properly understand the user's intent and to generate an appropriate response given the current dialog situation. In language \emph{generation} approaches, additional complexities may however arise, leading to common problems that the generated responses may be incomplete, grammatically or semantically incorrect, repeated, too general or too short \cite{song2018ensemble,JannachManzoor2020}.
Retrieval-based approaches are an alternative to such language generation approaches. The idea in such systems is to retrieve suitable responses from the dataset of recorded dialogs and adapt them to the context of the current dialog if needed. Retrieval-based approaches have been commonly applied in the domains of question-answering (Q\&A) systems and general purpose conversational systems \cite{riezler2007statistical,sakata2019faq,bonetta2021retrieval, banchs2012iris}. However, these approaches are not very common in the research literature for building CRS, where nowadays language generation models are predominant.

One potential advantage of retrieval-based approaches over generation-based ones comes from the fact that they return utterances originally made by humans and thus they are usually grammatically correct and in themselves semantically meaningful. Furthermore, retrieval-based approaches do not require the expensive training of complex models for language generation. On the other hand, retrieval-based methods might have deficits with respect to creativity and generalizability, i.e., the system may have problems to reasonably react to previously unseen situations.\footnote{See also \cite{yang2019hybrid} for a comparison of generation-based and retrieval-based methods.} Note, however, that a recent analysis of two neural language generation approaches in \cite{JannachManzoor2020} indicated that these systems almost exclusively return utterances that appeared in the same form in the training data as well, leading to a diminished advantage in this respect.

Ultimately, given the lack of recent research on building retrieval-based CRS, it is largely unclear how such systems would fare when compared to modern language generation approaches. With this work, we aim to close this research gap. Specifically, we propose a novel retrieval-based approach to building CRS and evaluate the quality of its responses with the help of a user study (N=90). As baselines in our comparison, we include two recent neural approaches named KBRD \cite{chen-etal-2019-towards} and KGSF \cite{zhou2020improving}, where KGSF is the more recent work and was reported to improve over KBRD in \cite{zhou2020improving}.

Overall, our study revealed that the study participants on average considered the responses returned by our retrieval-based system (named \crbCRS) to be of higher quality than those returned by the neural language generation approaches. This indicates that retrieval-based techniques should be considered as an alternative or complement when building a CRS. Furthermore, to our surprise and differently to what is reported in the literature, the more recent KGSF system did \emph{not} outperform the previous KBRD system in our human evaluation, which points to open questions regarding our evaluation approaches for CRS.

The rest of the paper is organized as follows. Next, in Section \ref{sec:previous-work}, we review selected works on generation-based approaches to CRS and provide a brief survey on retrieval-based approaches. Afterwards, in Section \ref{sec:system-details}, we present the technical details of our retrieval-based system. Section \ref{sec:evaluation-methodology} describes our research methodology and Section \ref{sec:results} presents the obtained results. Afterwards, we discuss additional insights we obtained from further exploratory analyses in the same section.
Finally, in Section \ref{sec:discussion}, we discuss the implications of our work.

\section{Previous Work}
\label{sec:previous-work}
In this section, we first revisit the details of KBRD and KGSF, the two recent \emph{language generation} approaches that we consider in our study. Next, we review \emph{retrieval-based approaches} that are typically used in Q\&A systems. Finally, we discuss questions regarding the evaluation of CRS.

\subsection{Language Generation Approaches: KBRD and KGSF}
Various neural approaches based on language generation models were proposed in recent years, which allow CRS to respond to user's utterances in natural language. Mainly, in these approaches, multiple neural network components including RNNs, GCNs, CNNs, GANs, encoder-decoders pairs etc.  were used to carry out the task of language generation and recommendation \cite{zhou2020improving, li2018towards,chen-etal-2019-towards,DBLP:journals/corr/JoshiMF17, kang2019recommendation, hayati2020inspired,yang2019hybrid}.
For example, in the work of Li et al.~\cite{li2018towards}, an RNN module is responsible for predicting the sentiment of user's utterance towards a given a preference (entity), and subsequently the outcomes of the RNN module are used as an input to the autonencoder-based module for computing the recommendation.

In this work, we focus on two recent language generation approaches to CRS, namely KBRD \cite{chen-etal-2019-towards} and KGSF \cite{zhou2020improving}, which were both developed and evaluated based on the \emph{ReDial} dataset mentioned above. The ReDial dataset was collected in the movies domain using crowdworkers for the development of a CRS named DeepCRS  \cite{li2018towards}. Specifically, the workers were asked to participate in a conversation, where one person played the role of a recommendation \emph{seeker} and the other was the (human) \emph{recommender}. Overall, the ReDial dataset consists of over 10,000 dialogs about movies.

The \emph{KBRD} system, published at EMNLP-IJCNLP '19, is technically based on a sequence-to-sequence encoder-decoder Transformer framework \cite{vaswani2017attention}. The authors mention that they preferred the Transformer framework over HRED \cite{Sordoni:CIKM2015} because of its often better performance in various NLP tasks, e.g., Q\&A or machine translation, see \cite{vaswani2017attention,ott2018scaling,liu2018generating,chen2019towards,rajpurkar2016squad, yang2018hotpotqa}. For the recommendation task, KBRD makes use of an additional external knowledge source based on DBpedia data \cite{lehmann2015dbpedia}. Thereby, the system utilizes information regarding entities (movies) and their features (e.g., `funny') mentioned by the seeker in the ongoing recommendation dialog.

The \emph{KGSF} system, presented at KDD '20, consists of multiple neural components. In addition to DBpedia, the KGSF system relies on another external knowledge graph (KG), named ConceptNet \cite{speer2017conceptnet}. Given a dialog utterance, the system first extracts word-oriented features, like semantic relations between two words, from ConceptNet and item-related features from DBpedia \cite{lehmann2015dbpedia}. Subsequently, to capture and encode the semantic relations between the previously extracted words, KGSF utilizes a Graph Convolutional Network (GCN) \cite{BMVC2016_114, DBLP:conf/iclr/KipfW17}. Similarly, to learn the representations of items and their features, the authors rely on a Relational Graph Convolutional (R-GCN) network  \cite{10.1007/978-3-319-93417-4_38}. Given the word and item node representations that were previously encoded, a Mutual Information Maximization technique \cite{48921} is used to improve and align the semantic representations. As a results of applying this technique,  fused KGs are produced that are subsequently used to generate the system responses.
Overall, the KBRD and KGSF systems share the similarity that they both rely on a Transformer framework and that they use the same (DBpedia) KG for the recommendation task.

\subsection{Retrieval-based Approaches}
Several retrieval-based approaches have been proposed in the literature for a variety of NLP tasks, e.g., machine translation \cite{ballesteros1997phrasal, elayeb2018towards}, Q\&A \cite{riezler2007statistical, sakata2019faq, rajpurkar2016squad}, or open-domain dialog systems \cite{8260795, sugiyama2013open, feng2019learning}. Recently, Q\&A systems obtained increased importance due to the fact that digital assistants such as Apple's Siri or Amazon's Alexa have become popular for answering day to day queries. However, such assistants often do not support multi-turn dialogs well, which is however an essential requirement of any CRS \cite{jannach2021crscsur}. From a technical point of view, various proposals from the literature can be related to retrieval-based conversational recommendation tasks.

\emph{AliMe} \cite{qiu2017alime} is a single-turn Q\&A system, which combines both retrieval and generation components. In this system, a sequence-to-sequence neural model is used to rerank a previously retrieved set of candidate responses. On similar lines, a retrieval-based Q\&A system is proposed in \cite{bilotti2007structured} that ranks the candidate responses using the query structure, annotations and keywords.

In addition to single-turn Q\&A systems, several hybrid approaches, i.e., ones that combine retrieval techniques with language generation,  were proposed in the literature for multi-turn \emph{dialog systems}. For instance, the authors of \cite{yang2019hybrid} present a hybrid solution, where a retrieval module returns a set of candidates from a dataset that contains queries and responses, and a generation module based on a \emph{seq2seq} model generates a response candidate given a conversation context. Similarly, an open domain hybrid dialog system is proposed in \cite{song2016two} in which a retrieval module returns a set of responses, and a \emph{biseq2seq} model \cite{zoph2016multi} generates a response. The output of both modules then undergoes a post-ranking step for the final selection of a response to a given user query. Another practical example of such a hybrid approach is Microsoft's \emph{XiaoIce}, a popular social chatbot system \cite{zhou2020design}, which determines candidate responses from two query-response databases.

Overall, while such hybrid approaches were found to be useful for Q\&A-related and general dialog systems, we did not find works that approach the conversational recommendation problem in a similar way, i.e., by coupling the outputs of both language generation and retrieval modules. Furthermore, we are not aware of any natural language based CRS that aims to retrieve responses from the recorded dialog dataset as we do in our present work. In the future, we however plan to combine retrieval and generation-based approaches for building  CRS.

\subsection{Evaluation of Conversational Recommender Systems}
Generally, a CRS can be evaluated in various quality dimensions. According to \cite{jannach2021crscsur}, we can differentiate between system effectiveness, efficiency, conversation quality and ``effectiveness of subtask''. System \emph{effectiveness} is mainly related to the question if a CRS is able to help users finding relevant information or making better decisions. \emph{Efficiency} indicates how much effort is required by the user. \emph{Conversation quality} may cover aspects like the fluency or understandability of the system responses. \emph{Effectiveness of subtask} finally refers to the system's performance for specific subtasks like intent recognition or entity recognition.

From a methodological point of view, quality measurements are typically either made with the help of computational (``offline'') experiments or with studies involving humans.\footnote{Other forms of evaluations, including field tests (A/B) tests or exploratory research methods, are relatively rare in the literature on CRS.} In offline experiments, system effectiveness is often measured in terms of \emph{recommendation quality}, and metrics like precision or recall are used as proxies in such evaluations. This form of evaluation is also the predominant evaluation approach for non-conversational recommender systems. The authors of the KBRD and KGSF systems discussed in this present work were also evaluated with the help of such measures. In case of KBRD, the system was evaluated using recall, and it turned out that it was favorable over the previous DeepCRS system \cite{li2018towards}. In the case of KGSF, recall was used as well, this time using KBRD and DeepCRS as baselines that were outperformed both in normal and in cold-start situations.

In addition to recommendation quality (system effectiveness), the authors of KBRD and KGSF also used offline experiments to assess the quality of the conversations. Specifically, they used linguistic measures such as \emph{distinct n-gram} and \emph{perplexity} to assess the diversity and fluency of the system-generated responses. Looking at the study outcomes, the authors of KBRD found their system to be favorable over DeepCRS also in these dimensions. For the KGSF approach that was published later, the results were even better according to the authors.

While offline experiments are common in the literature, they have various limitations. It is for example not always clear to what extent metrics such as precision and recall actually correspond to the \emph{perceived quality} of a system. Also for linguistic measures there are open questions regarding the representativeness of such measures for human perception in certain cases \cite{liu-etal-2016-evaluate}. Moreover, as discussed in \cite{JannachManzoor2020}, neither the DeepCRS nor the KBRD system actually generate new sentences to a large extent. The linguistic scores therefore mainly measure the quality of utterances that are genuinely human.

Studies involving humans are therefore not uncommon in the CRS literature, and such studies are commonly applied to gauge all sorts of quality factors related both to system effectiveness (e.g., perceived recommendation quality), efficiency (e.g., perceived effort), and conversation quality (e.g., perceived naturalness of the system responses). The authors of KBRD and KGSF also relied on an additional study involving humans when evaluating their systems. Specifically, in both cases human judges were tasked to assess the quality of different system responses in a set of dialog situations. In case of the KBRD system, the judges had to score the responses by the proposed approach and the DeepCRS baseline system on a 0-2 scale in terms of \emph{consistency} with the previous dialog. A similar approach was used for the KGSF system. This time, however, the judges were asked to asses the \emph{fluency} and \emph{informativeness} of the system responses in a given dialog situation.  The KBRD and DeepCRS systems were used as baselines in this evaluation.

Both in the case of KBRD and KGSF, little is said about the background of the human evaluators and the specifics of the study setup. It therefore remains unclear if the judges were able to reliably and consistently assess concepts like consistency, fluency, or informativeness. Also, the number of judges was limited in both cases, e.g., there were three people involved in the human evaluation of the KGSF system and no information is provided about their linguistic competence.

Given the potential limitations of offline experiments, we rely on a study with humans in our work. Differently from the discussed previous works, we involve a much larger set of participants, and we do not require them to have any linguistic background or skills. Correspondingly, we also use a more general concept of ``(perceived) meaningfulness'' of the system responses as an evaluation criterion, which can refer to different aspects such as recommendation quality or the appropriateness of the response in a given dialog situation.

\section{A Contextual Retrieval-based CRS (\crbCRS)}
\label{sec:system-details}
\begin{figure} [ht]
  \centering
  \includegraphics[width=1.0\textwidth]{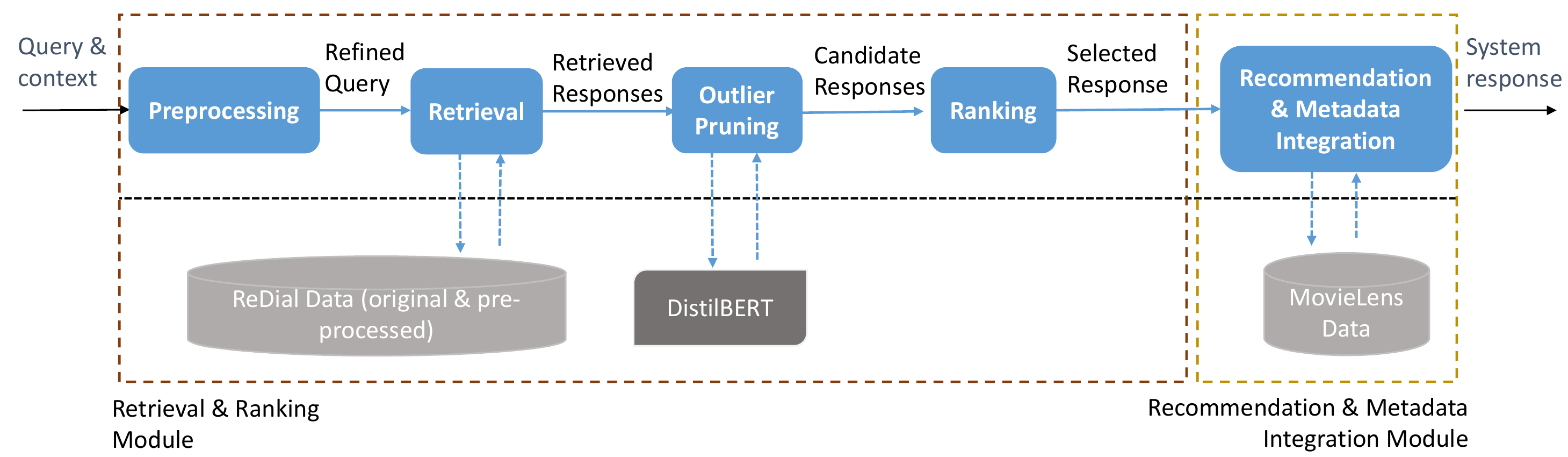}
  \caption{Overall Architecture of the \crbCRS System}
  \label{fig:architecture}
\end{figure}

The contextual retrieval-based conversational recommender system (\crbCRS) proposed in this work is developed based on the assumption that suitable responses to user's utterances are contained in the recorded conversational recommendation dataset. 
In principle, our approach is not limited to a particular domain or dataset. In our case, however, we used the ReDial dataset for the movies domain to be able to compare it with previous systems. Overall, the architecture of \crbCRS, as shown in Figure \ref{fig:architecture},  has two main components: (1) a Retrieval \& Ranking Module, and (2) a Recommendation \& Metadata Integration Module.

\subsection{Retrieval \& Ranking Module}
\label{sec:retrieval}
Given the last user utterance and the history of the ongoing dialog, the \emph{Retrieval \& Ranking} module first finds a set of candidate responses from the dataset, which are then filtered and ranked according to their assumed fit. The best-ranked candidate is finally forwarded to the next module.

\paragraph{Response Retrieval} Given a seeker utterance and the history of the dialog, this module, in a first step, determines the
most similar \emph{seeker} utterances in the ReDial datatset based on similarity scores. To compute these similarity scores, we prepared the ReDial dataset in advance by applying common preprocessing techniques such as conversion to lowercase or removal of stop words, and by replacing movie IDs with a placeholder. We then computed a \emph{TF-IDF} encoding\footnote{We also tried a semantic encoding with Sentence-BERT \cite{reimers2019sentence}, but this did not lead to better results.} of the utterances in the preprocessed dataset, and we use the cosine similarity measure to retrieve the most
similar seeker utterances from the dataset.
The candidate responses in our approach are  those sentences in the ReDial dataset that immediately follow after the retrieved similar seeker utterances.

Note that considering only the very last seeker utterance in an ongoing dialog may be too restrictive, as this last utterance may not convey the full context of the conversation. We therefore compute multiple sets of candidate responses, where each set is computed by providing the system with more context, i.e., by increasingly providing more previous utterances from the dialog history for the retrieval task. With this approach, our goal is to obtain responses that are structurally more diverse and more coherent with the dialog history. In the context of this work, we considered four configurations when retrieving similar seeker utterances\footnote{Alternative configurations are possible, e.g., by considering more elements of the ongoing dialog or by considering only certain types of utterances.}:

\begin{enumerate}
    \item Using only the last seeker utterance.
    \item Using the last seeker utterance along with the previous human recommender response from the ongoing dialog.
    \item Using the last seeker utterance along with the previous pair of seeker-recommender utterances.
    \item Using the complete dialog history until the last seeker utterance.
\end{enumerate}

\begin{figure} [h!t]
  \centering
  \includegraphics[width=0.9\textwidth]{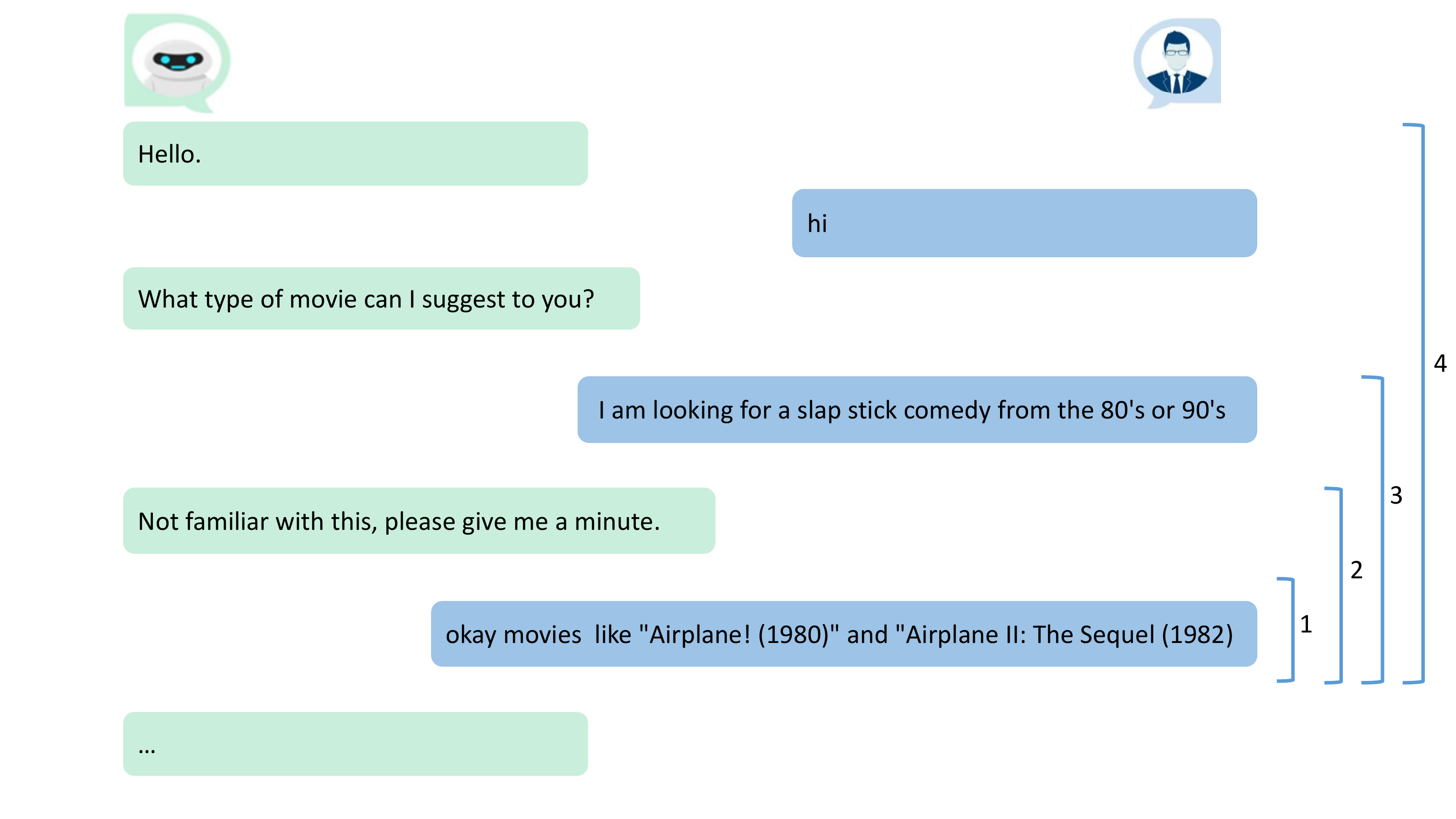}
 \caption{Dialog Context Illustration}
  \label{fig:dialog-history}
\end{figure}

An illustration of this approach of incrementally extending the conversation context is shown in Figure \ref{fig:dialog-history}. Technically, we combine the relevant utterances of the ongoing dialog when determining the similarity scores for each situation.

As a result of this process, we create up to four sets of candidate responses, depending on the length of the dialog history. In each of these sets, we retain the \emph{n} most similar seeker utterances, where \emph{n} is a system parameter. In our experiments, we used \emph{n=5} according to the results of a pre-test.
Therefore, using these settings, we may have up to 20 different similar seeker responses, and thus candidate responses, at this stage (up to 4 sets with 5 elements each).

As a first measure to assure the quality of the system responses, we applied a filtering step, where we try to avoid too long or too short candidates. The underlying assumption in this step is that too short responses might not contain enough relevant information, and too long responses often refer to more than one intent. In our experiments, we therefore removed all candidates which contained less then 3 words or more than 20 words. These thresholds were based on the average length of 9.7 words (SD = 7.8) of the human recommender responses in the ReDial dataset. With the chosen thresholds, we considered about 75\% of the recommender responses in our retrieval process. The overall retrieval approach is presented in Algorithm 1.

\begin{algorithm}
\label{alg:response-retrieval}
	\caption{Response Retrieval Approach (sketch)}
		\hspace*{\algorithmicindent} \textbf{Inputs:} \\
		\hspace*{\algorithmicindent} q $ \gets$ User's last utterance (may include parts of the dialog history)\\
		\hspace*{\algorithmicindent} D $\gets$ {Original ReDial dataset, with indexed utterances} \\
		\hspace*{\algorithmicindent} M $\gets$ {TF-IDF encodings of utterances in D, using same index} \\
		\hspace*{\algorithmicindent} j $\gets$ {3 (tuned value), lower limit for words in candidate response} \\
		\hspace*{\algorithmicindent} k $\gets$ {20 (tuned value), upper limit for words in candidate response} \\
		\hspace*{\algorithmicindent} $n \gets$ {5 (tuned value), number of responses to be retrieved} \\
	\hspace*{\algorithmicindent} \textbf{Output:} $S \gets$ {Set of retrieved candidate responses}
	\algrule

	\begin{algorithmic}[1]
	    \State $counter \gets$  0
		\State $q' \gets$ Preprocess and tokenize q
		\State $q_v \gets$ Compute TF-IDF representation of q'
		\State $S_m \gets$ Cosine similarity between $q_v$ and all seeker utterances in M, list of values, indexed like M.
		\State $S_{ranked} \gets$ Create ranked list of of seeker utterances according to similarity scores in $S_m$, in descending order
		\For {\emph{utterance} in $S_{ranked}$}
		\State $s_i \gets$ Index of \emph{utterance} in $D$
		\State $r \gets  D[s_i+1] $ \% Return immediate response to utterance from ReDial
		\If {length($r$) $\ge$  j AND length($r$) $\leq$ k}
		\State $S \gets$ Add $r$ to  $S$
		\State $counter$++
		\If{$counter$ = $n$}
		\State \Return S
		\EndIf
		\EndIf
		\EndFor
	\end{algorithmic}
\end{algorithm}

\paragraph{Outlier Pruning}

In the next step, our goal was to see if there is some dominant ``theme'' in the candidate responses. The underlying intuition is as follows. If, for example, a large majority of the candidate responses contain a movie recommendation, and only a small set of candidates are chit-chat, candidate responses that contain a recommendation should be ranked higher. Technically, we implemented this intuition by computing, for each of the four candidate sets, the pairwise similarity of the candidates and by then retaining only those candidates that are highly similar to other ones.

Note that natural recommendation conversations may contain a substantial portion of general or open domain utterances for example chit-chat utterances, see also \cite{cai2019department,kang2017understanding}. Therefore, to compute the similarities of the responses, we used an open-domain language model for this purpose instead of the relatively limited ReDial dataset.
Specifically, we relied on a pre-trained lighter version of BERT called DistilBERT\footnote{\url{https://huggingface.co/distilbert-base-uncased}}. DistilBERT \cite{Sanh2019DistilBERTAD} is a Transformer-based model trained on textual data from a wide variety of sources and has proven to lead to solid results in open domain natural language tasks like question-answering or text classification \cite{mozafari2020method, liu2021improving}.

Technically, given a set of candidates, we compute similarity scores between each pair of candidates using \emph{DistilBERT} embeddings. Subsequently, we retain only one-fourth\footnote{This threshold is again a system parameter to be determined in pre-tests. Note that we conducted pre-tests with another set of randomly sampled dialogs from the test set to select suitable parameters and to fine-tune the approach.} of the most similar response pairs in the resulting set, i.e., we discard less similar pairs and retain only a smaller number of candidates.  In our specific setting, all four sets of candidate responses undergo this pruning process. As a result, only structurally similar responses are contained in each set, and outliers should have been removed. This outlier pruning process, see Algorithm \ref{alg:outlier-pruning}, technically leads to a loss of information compared to the original set of responses. However, according to our experiments, it helps us retain the most relevant information and thus most appropriate response candidates.

\begin{algorithm}
	\caption{Outlier Pruning Technique for a Set of Candidates}
    \label{alg:outlier-pruning}
		\hspace*{\algorithmicindent} \textbf{Input:} S $\gets$ Non-empty set of retrieved candidate responses \\
	\hspace*{\algorithmicindent} \textbf{Output:} $R \gets$ {Pruned set of mutually similar responses}
	\algrule
	\begin{algorithmic}[1]
		\State  $P$ $\gets$  {Create set of all pairs of elements of S}
	    \State $t \gets Floor(size(P)/size(S)) $ \% Heuristic for threshold regarding pairs to retain
	    \For {candidate pair p ($c_1$, $c_2$) $\in P$}
	    \State  $v_1 \gets { BertVectorizer(c_1)}$ \% Calculate BERT embedding
	    \State  $v_2 \gets { BertVectorizer(c_2)}$
	    \State $cosine_p \gets{ Cosine(v_1, v_2)}$ \% Compute cosine similarity
        \State Add tuple $(p,cosine_p)$ to a list $L$
	    \EndFor
        \State Sort $L$ based on $cosine_p$ values, in descending order
        \State $R \gets $ top-\emph{t} pairs of candidates from $L$
        \State \Return $R$
	
	\end{algorithmic}
\end{algorithm}

\paragraph{Response Ranking}
The final goal in this module is to rank the candidate responses and select the most plausible response for a given user utterance and dialog context.

As a first step, we merge the candidate sets and remove duplicate responses. Next, our goal is to give more weight to candidate responses that are assumed to be fluent and naturally structured. We accomplish this by computing a perplexity score $r$ for each candidate as a proxy for fluency, see also \cite{chen-etal-2019-towards,deshmukh2010role}.
Technically, we first compute the likelihood probability of neighboring bigrams over the human-recommender responses in the underlying dataset. As proposed in \cite{song2018ensemble}, using bigram probabilities in the underlying data represents a logical step to determine the internal structure of the responses within the co-occurrence of neighboring words. Note that only the sum of these individual probabilities is sensitive to the length of the particular response. Therefore, following \cite{song2018ensemble}, we calculate the average of these individual probabilities by taking the ratio of the sum of probabilities and the total number of bigrams to obtain a final perplexity score ($\mathcal{F}_{r}$) for a particular candidate response $r$.  Finally, the response with the highest score ($r^*$) among the candidates may be selected as a system response.
The ranking approach is formally described in Equation \ref{eq:perplexity-scoring}.

\begin{equation}
r^* = \argmax_r \mathcal{F}_{r}
\left( \frac{\sum_{i=0}^{k} log \frac{|\mathcal{B}(w_i, w_{i+1})|}{ |\mathcal{U}_{wi}|+|\mathcal{U}|}}
{|r_\mathcal{B}|}\right), \forall  r \in \mathcal{R}
\label{eq:perplexity-scoring}
\end{equation}

Here, $|\mathcal{B}(w_i, w_{i+1})|$ denotes the number of occurrences for neighboring bigrams, $|\mathcal{U}_{wi}|$ represents the number of occurrences for the given word $w_i$, $\mathcal{U}$ is the set of unigrams in the dataset, and  $|r_\mathcal{B}|$ and $k$ denote the number of bigrams and unigrams, respectively, for a given candidate response $r$ from the set of candidate responses $\mathcal{R}$.

While simply returning the candidate response with the highest perplexity score at this step leads to good results, we found that further fine-tuning the ranking can help to improve the performance of our system. Specifically, we used a relatively simple heuristic regarding potential seeker intents and their match with the candidate responses. In pre-tests we often found that whenever the last seeker utterance contains a reference to a movie name, it is a good strategy to answer with a recommendation (compared, e.g., to a chit-chat response). Likewise, if we think that the last seeker utterance is a chit-chat, e.g., ``\emph{Thank you for the recommendations}'', we should probably not make a recommendation and rather respond with a chit-chat answer.  Technically, we implemented this intuition by simply adding a positive integer weight to the perplexity scores depending on the situation, and by then re-ranking the candidates after this weight update before returning the candidate response with the highest weight. In our experiments, we determined suitable values for these weights empirically, considering also the distribution of values for the perplexity scores. We observed that the obtained perplexity scores range from -1 to -5. Therefore, to reliably push the preferred responses, we added a weight of +5 to prioritize certain candidates with a recommendation. In the case of chit-chat responses, a weight factor of +2 was added.

To discriminate between chit-chat and other utterances, we used a simple heuristic decision rule. Based on previous observations in \cite{JannachManzoor2020}, we identified a small set of keywords (e.g., `thanks' or `bye'), which we considered indicative of a chit-chat utterance. More elaborate approaches are possible as well.

\subsection{Recommendation and Meta-data Integration Module}
At this stage, our system has determined one single best response candidate from the underlying dataset. Remember, however, that the retrieval was based on a pre-processed dataset in which we replaced movie IDs with placeholders. Therefore, before finally returning the response, we fill the placeholders, in case such placeholder exist in the response, with movie information that suits the assumed preferences of the human seeker in the ongoing dialog.

The \emph{Recommendation \& Meta-data Integration} module of our system implements two recommendation strategies, supporting two specific intents.

\begin{enumerate}[label=\emph{(\roman*)}]
  \item In case the seeker has mentioned movies in the ongoing dialog, we assume the user is seeking a recommendation by mentioning a movie that he/she likes\footnote{We interpret any movie mention as a positive preference here. More complex approaches, e.g., based on sentiment analysis, are possible, see \cite{li2018towards}.}, as done in \cite{zhou2020improving,chen-etal-2019-towards}.
  \item Otherwise, in case any of about 30 predefined genre keywords\footnote{These keywords can be derived from the MovieLens dataset.} is detected in the last seeker utterance or the previous dialog history, we assume that the user is looking for movies of a certain genre.
\end{enumerate}

Depending on the situation, one of two recommendation strategies is applied. In the first case, movie-based recommendations \emph{(i)}, our system takes the last mentioned movie in the dialog and returns a movie that is highly similar to it. To this purpose, we once apply matrix factorization on the MovieLens-25M dataset. Technically, we made use of Python's \emph{sklearn} library using the \emph{TruncatedSVD} method with the number of latent factors set to 20. Then, given the last mentioned movie, we return as a recommendation that movie from the MovieLens dataset which (a) overlaps with it in terms of the genre and (b) is the closest to it in the latent space. Again, we use cosine similarity as a distance measure between the item embeddings. In the case of genre keyword-based recommendation \emph{(ii)}, we match the last mentioned genre in the dialog history with the genres of the movies in the dataset using the cosine similarity measure and we rank the items based on genre similarity. Finally, in both cases \emph{(i)} and \emph{(ii)}, we apply additional popularity filters to avoid too obscure recommendations, which can appear in purely similarity-based approaches, see also \cite{Ekstrand:2014:UPD:2645710.2645737}. Specifically, we apply empirically determined minimum thresholds for the popularity of the recommendable movies in terms of their mean ratings and we avoid the recommendation of very old and probably less known movies.

Clearly, for both recommendation approaches, more sophisticated strategies are possible. Furthermore, note that in theory it can happen that we have a placeholder in the response, but no movie or genre mention in the dialog history. In this case, which did not happen in any of our experiments, we would take that movie as a recommendation that was mentioned in the original utterance. In case a response consists of both a genre keyword and a movie mention, we apply the movie-based recommendation approach.

Finally, before the system returns the response, we replace the movie title in the candidate response with the title of the movie determined with one of the described strategies. Moreover, we also replace relevant metadata in the candidate response. For that purpose, we rely on a database containing meta-data for the movies used in our previous research \cite{jannachmultirow2021}. This metadata replacement is done based on a small set of heuristics. For example, if the selected response is ``Another funny movie is [Movie Title]", we replace any genre keyword (e.g., here ``drama'') with the actual genre for the recommended movie. Similarly, we also do replacements for the mentions of actors based on a small set of string-matching rules. In some cases, when we detect that the seeker requests description of a movie, we return a plot summary from the additional meta-data mentioned above.

In that context, our system is similar to KBRD and KGSF, which both rely on external metadata during the recommendation process. For our system, please note that the string-matching rules at this stage were \emph{not} designed based on the 70 dialog situations that we used in the experiment that we describe next. Instead, we created another random set of 70 dialog situations before the experiment, which we analyzed when crafting the meta-data integration rules.

\section{Evaluation Methodology}
\label{sec:evaluation-methodology}

We conducted an online user study to compare the perceived quality of our retrieval-based CRS with the perceived quality of the language generation approaches KBRD and KGSF. To be able to obtain a fine-grained assessment, the specific task of the study participants was to assess the \emph{quality of individual responses} in a given dialog situation, as was done previously in \cite{zhou2020improving,li2018towards,chen-etal-2019-towards}.

\begin{figure} [h!t]
  \centering
  \includegraphics[width=0.9\textwidth]{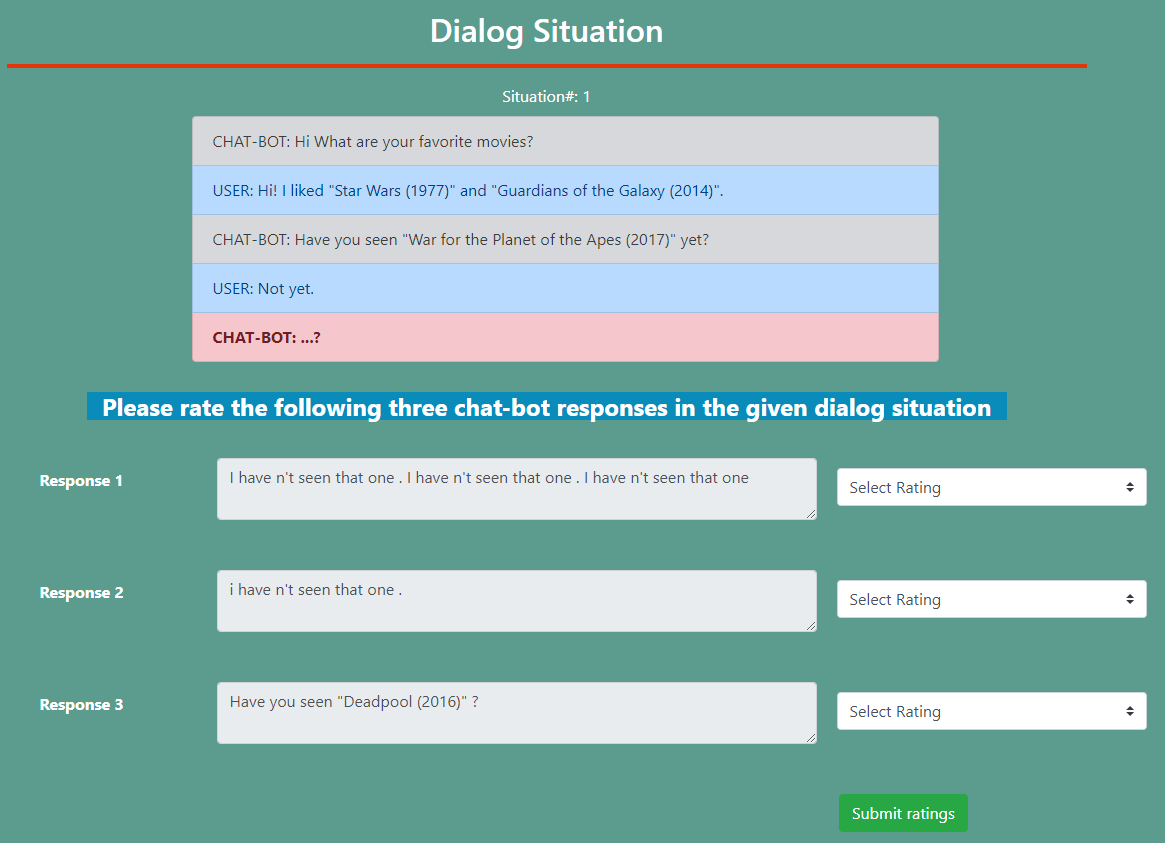}
 \caption{Response Rating User Interface}
  \label{fig:user-study}
\end{figure}

\paragraph{Experiment Settings}
To conduct the experiment, we reused the code provided by the authors of KBRD and KGSF. To generate the responses given the last seeker utterance, we trained the models by ourselves using the same hyperparameter settings and data splits that were used in the original works \cite{zhou2020improving, chen-etal-2019-towards}\footnote{This is appropriate as we used the same data and data splits and assume that the authors tuned the hyperparameters appropriately for their works.}.
For \crbCRS, we split the ReDial dataset into training and test sets as was done for KBRD and KGSF. Note that all experiments presented and discussed in this work, including the original ones, were conducted in the same domain, that of \emph{movie} recommendation.
\paragraph{Methodology}
The study procedure was as follows. The participants interacted with a web application\footnote{\url{https://study-ainf.aau.at/}} that was developed for the purpose of the study. After informed consent, the participants were presented with a dialog situation that ended with a \emph{user (seeker)} utterance, as shown in Figure \ref{fig:user-study}. Below the dialog situation, we showed participants three possible system responses to the last seeker utterance. Each of the responses was created with the help of one of the compared systems in our study (KBRD, KGSF, \crbCRS).

The order of the responses was randomized. The task of the study participants was now to assess the quality of each response in terms of its meaningfulness using a 5-point scale. The provided scale ranged from ``Entirely meaningless'' to ``Perfectly meaningful''. At the beginning of the study, instructions were provided to the participants how they should evaluate the meaningfulness of a response. For example, a response should be a logical continuation of the dialog. Furthermore, in case the response included a recommendation, the participants were instructed to assess if the recommendation seems meaningful given the seeker-recommender interactions in the dialog so far. In the course of the experiment, each participant had to evaluate ten dialog situations. After the participants had scored the responses for ten dialog situations, users were asked a number of questions related to their demographics. We also implicitly recorded the total time a participant spent on the study. We share all study material and the source code used to build \crbCRS online\footnote{\url{https://github.com/ahtsham58/CRB-CRS}} to ensure reproducibility of our work.

The dialog situations in the study were created as follows. First, we randomly selected 70 dialogs from the ReDial dataset\footnote{These dialogs were not part of the dataset that we used to compute TF-IDF vectors in Section \ref{sec:retrieval}.}. Then, for each dialog we again randomly removed a part of the tail of the dialog, making however sure that the resulting dialog situation ends with a seeker utterance. A dialog situation is therefore always a subset of the original dialog that starts at the beginning and ends with a seeker utterance. By randomly choosing the situation, where the recommender should respond, we ensure that we cover dialog situations at various stages, e.g., dialogs that have just stared, dialogs that are almost done, or dialogs where an active exchange between seeker and recommender is happening.

We recruited N=107 participants through the crowdsourcing platform Amazon Mechanical Turk (AMT). To ensure the reliability of the obtained results, we enforced that study participants are fluent in English and have some interest in movies. Moreover, to detect cases where crowdworkers did not complete the task with the necessary care, we used one of the ten dialog situations for an attention check. Specifically, in one of the 10 dialog situations we asked the participants to select a particular rating from the given scale. We considered the attention check to be failed whenever a participant did not select the required score, and we removed all data from such unreliable crowdworkers.

\section{Results}
\label{sec:results}
Here, we first present the main outcomes of our study (Section \ref{subsec:main-results}), before we discuss the observations from additional analyses (Section \ref{subsec:additional-analyses}).

\subsection{Main Results}
\label{subsec:main-results}
From the 107 participants that were recruited through AMT, 98 (about 92\%) passed the attention check. Furthermore, to further improve the reliability of the collected opinions, we manually inspected the provided scores for dubious patterns that might demonstrate a lack of attention by the participants. For example, one problematic pattern we came across was that sometimes a CRS responded with a recommendation for a movie which it already recommended earlier in the same dialog. Study participants who only looked at the last seeker utterance but ignored the dialog history would miss such cases of non-meaningful system responses. To avoid cherry-picking of such responses, we completely removed participants who demonstrated such patterns. We also provide the specific rules for removing participants and document all cases that we eliminated that way in the online material.

After this process, we ended up with 90 participants for which we are highly confident that they attentively completed the tasks. The summary statistics regarding the considered and discarded cases are shown in Table \ref{tab:statistics-cases}.

\begin{table}[h!t]
\renewcommand\thetable{1}
\vspace{6pt}
  \centering
  \begin{tabularx}{10cm}{lc}
  & \textbf{Total}\\ \toprule
  Number of participants (cases)  & 107 \\
  Number of participants passing the attention check &  98 \\
  Number of manually discarded cases &  8 \\
  Number of valid cases at end &  90 \\
  \bottomrule
  \end{tabularx}
  \caption{Statistics about Participants and Discarded Cases}
  \label{tab:statistics-cases}
\end{table}

The demographics of the participants are presented in Table \ref{tab:demographics}. The majority of the participants are fluent in English and frequent movie watchers. This indicates that the participants on average had suitable skills and background knowledge to accomplish the task. The overall average task completion time for the participants was about 9 minutes and we paid them 1.5 USD for their work.

\begin{table} [h!t]
\renewcommand\thetable{2}
  \centering
\small
\begin{tabular}{p{4.2cm}p{3cm}ccc}
\hline
 \textbf{Demographic Feature} & \textbf{Scale} & \textbf{Total} \\
\hline
\multirow{3}{4.2cm}{Gender}	          & Male  & 45 \\
                  & Female &  44 \\
                  & Other &  1
                  \\\hline
\multirow{5}{4.2cm}{Age}               & 18-25 &  6 \\
                  & 25-30 &  26\\
                  & 30-35 &  20\\
                  & 35-45 &  21\\
                  & 45-100 &  17
                   \\\hline
\multirow{4}{4.2cm}{English fluency level} & Beginner &  0  \\
                      & Intermediate & 0\\
                      & Fluent &  87\\
                      & Advanced &  3
                      \\\hline

\multirow{5}{4.2cm}{Education level} & High school or less & 13  \\
                      & Bachelor's &  51\\
                      & Master's &  22\\
                      & Doctorate & 2\\
                      & Other &  2
                      \\\hline
\multirow{5}{4.2cm}{Movie watching frequency} & Everyday &  17 \\
                      & Several times a week &  39 \\
                      & Once in a week &  21\\
                      & Once every few weeks & 9\\
                      & Less frequent &  4
                      \\\hline

\end{tabular}
\caption{Demographic Details of Participants}
\label{tab:demographics}
\normalsize
\end{table}

After the experiment and data processing, we had valid feedback for 810 dialog situations, i.e., 9 per participant. Remember that one of the 10 dialog situations was an attention check. The average scores and standard deviations for the each system are presented in Table \ref{tab:results-study-1a}. On average, our retrieval-based \crbCRS system outperformed both language generation approaches. Furthermore, the standard deviation is the lowest for this approach.

\begin{table}[h!t]
\renewcommand\thetable{3}
\centering
\begin{tabular}{lccc}
    & \textbf{KGSF} & \textbf{KBRD} & \textbf{\crbCRS} \\  \hline
 Average score  & 3.26 & 3.62  & \textbf{3.78} \\ \hline
 Std. deviation  & 1.48 & 1.33 & \textbf{1.21} \\
  \hline
\end{tabular}
\caption{Results of Scoring of System Responses}
\label{tab:results-study-1a}
\end{table}

To analyze the significance of the differences in the mean scores, an ANOVA analysis was conducted, which exhibited that the differences between the mean scores of the three groups are statistically significant ($p <0.01$). A Tukey HSD test further revealed that the differences in all pairwise comparisons between the three systems are also significant, with $p < 0.05$ in all cases.

\begin{figure} [h!t]
  \centering
  \includegraphics[width=1.0\textwidth]{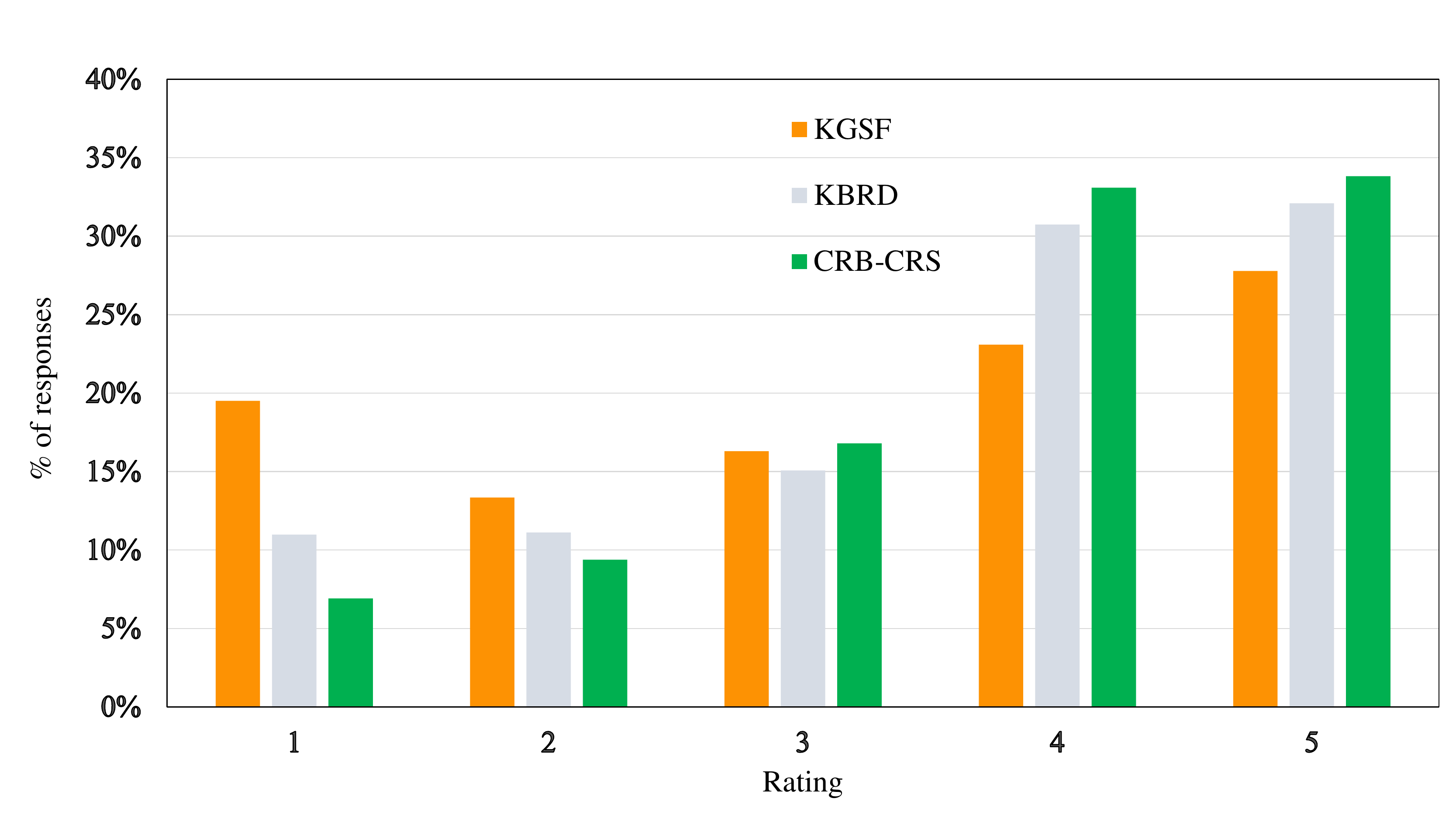}
 \caption{Distribution of Scores}
  \label{fig:ratings}
\end{figure}

A histogram that visualizes the distribution of the ratings with respect to the responses is shown in Figure \ref{fig:ratings}. It is evident that \crbCRS made the fewest mistakes, and similarly, \crbCRS received very good ratings more frequently than KGSF and KBRD. To our surprise, the KGSF system performed worse compared to KBRD, which is not in line with the results of the study reported in \cite{zhou2020improving}, where KGSF outperformed KBRD. Remember, however, that the human judges in \cite{zhou2020improving} evaluated the system responses in terms of fluency and informativeness, whereas we use the general concept ``meaningfulness''.

\subsection{Additional Analyses}
\label{subsec:additional-analyses}
Next, we report our observations from three additional analyses that we made to obtain a better understanding of the experiment outcomes and the underlying problem setting.

\subsubsection{Analysis of Failure Situations: Specific Seeker Queries}
\label{subsec:failure-specific-question}
A manual inspection of the collected data indicated that there are certain situations, where all compared systems struggle to return meaningful responses. In particular, these are often situations where the \emph{seeker} asks very specific questions, e.g., about the genre of a movie, if a certain actor is starring in a certain movie, or about the recommender's opinion about a movie. To understand the extent of the problem and if there are differences between the compared systems, we studied the system responses in such cases in more depth.

Looking at the valid cases that were collected in the study, we identified 96 instances, where the given dialog situation ended with a specific seeker question. In total, there were 34 unique dialog situations.\footnote{Remember that a particular dialog situation may be presented to more than one study participant. The identification of relevant dialog situations was done in a manual process.} The average scores and standard deviations for both unique and total cases are shown in Table \ref{tab:results-Study-1c}.

\begin{table} [h!t]
\renewcommand\thetable{5}
  \centering
\small
\begin{tabular}{p{2cm}p{2.5cm}ccc}
\hline
 \textbf{} & & \textbf{KGSF} & \textbf{KBRD} & \textbf{\crbCRS}  \\
\hline
\multirow{2}{2.5cm}{Unique cases} & Average score & 2.61 & 3.17 & \textbf{3.74}  \\
                  & Std. deviation & 1.40 & 1.47 & \textbf{1.14}
                  \\\hline
\multirow{2}{2.5cm}{Total cases}& Average score & 2.76 & 3.29 &\textbf{3.94} \\
                  & Std. deviation & 1.42 & 1.51 &\textbf{1.03}
                   \\\hline
\end{tabular}
\caption{Results of Response Scores for Specific Seeker Questions}
\label{tab:results-Study-1c}
\normalsize
\end{table}

\begin{figure} [h!t]
  \centering
  \includegraphics[width=1.0\textwidth]{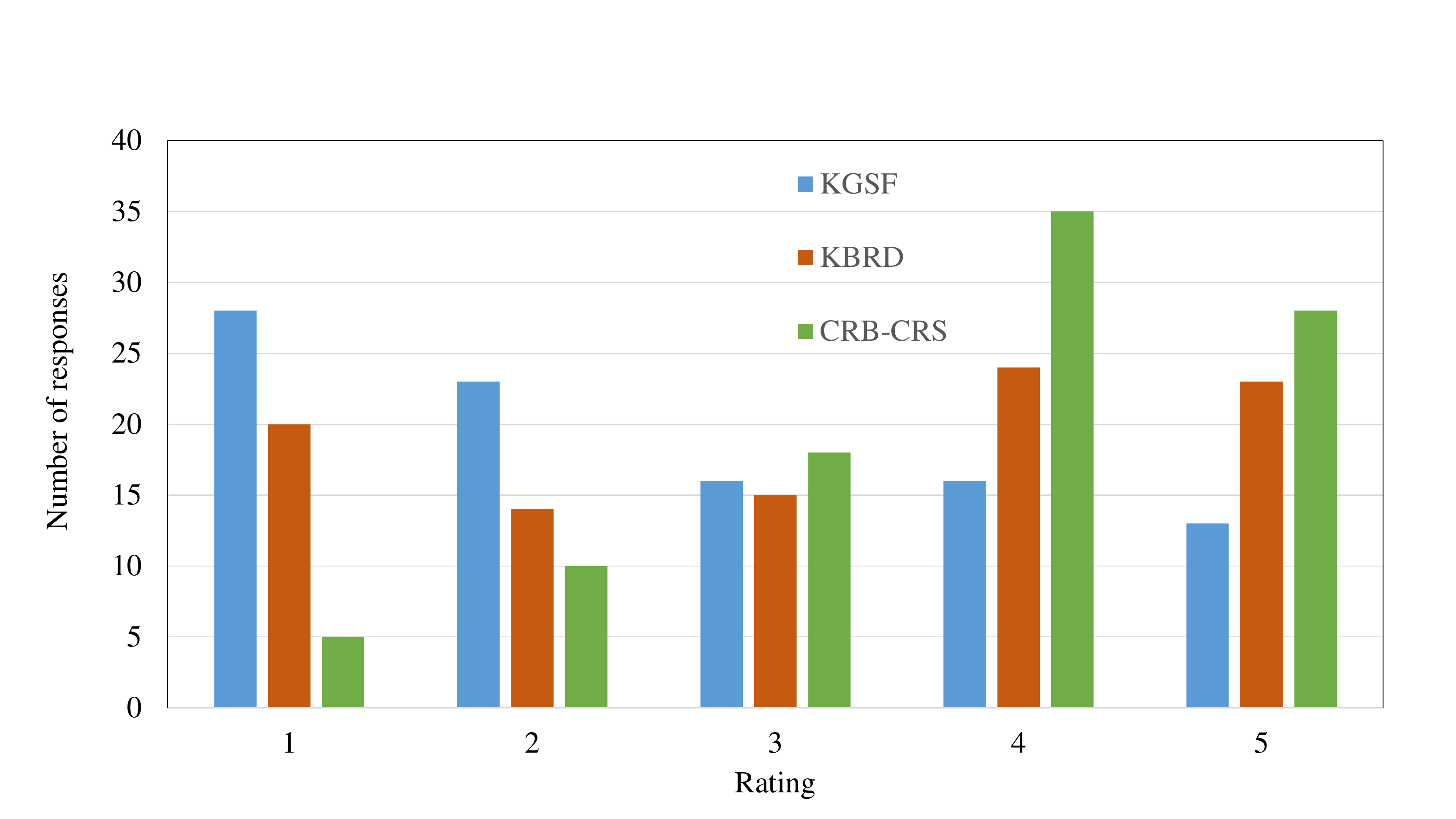}
 \caption{Distribution of Scores (Responses to Specific Seeker Questions)}
  \label{fig:ratings-questions}
\end{figure}

Again, our \crbCRS system outperformed KGSF and KBRD, and the standard deviations are the lowest as well in this comparison. A histogram in Figure \ref{fig:ratings-questions} visualizes the distribution of scores for all cases shows that the differences mainly stem from the extreme values, where \crbCRS more often obtained very good scores and infrequently responded to the seeker's questions in a non-meaningful way that would have lead to low scores. One reason for this phenomenon could be the exploitation of meta-data information in the final step of our system before the responses are returned.

Overall, it appears that all compared systems may have difficulties to respond to specific seeker questions. Looking closer at individual cases, we observe two very common types of problems. First, in the cases when a seeker asks for the metadata information, e.g., about actors or the genre, regarding a recommended movie, the systems often either provide entirely wrong metadata or return completely unsuited responses. Second, we found a number of cases in which especially the language generation systems respond with a very general or evasive answer, e.g., ``\emph{I haven't seen that one yet}''. Our \crbCRS system, in contrast, more often responds with a movie recommendation than with a general response. What we sometimes observe is that study participants seem to be more lenient in such cases, where the system made a recommendation even though a specific question was asked. At least, it seems more acceptable to users than a factually wrong or disconnected answer.

Generally, all systems might also suffer from the characteristics of the underlying ReDial dataset. The recorded conversations in this dataset are often ``instance-based'', i.e., the dialog partners rather talk about specific movies than about genres or metadata. In parts, this may have to do with how the data was created with the help of crowdworkers, who were instructed to create dialogs that contain a specified minimum number of movie mentions. As a result, only a smaller set of utterances is related to meta-data. Also, there are very few cases, where a human seeker was asking for an explanation for a provided recommendation.

\subsubsection{An Exploratory Analysis of Challenging Dialog States / User Intents}
After our analysis of the frequent failure situation when seekers ask specific questions, our aim was to better understand which other dialog situations may be challenging for all systems. In particular, we tried to identify which user \emph{intents} are difficult to handle, see \cite{CaiChen2020UmapIntents} for a study on user intents identified in the ReDial dataset.

We conducted an exploratory analysis of the collected study data to obtain a first impression where all systems face problems. To that purpose, we manually inspected the valid dialog situations from the user study and looked for cases where all three system responses received the lowest or second-lowest possible score (``entirely meaningless'', ``meaningless''). We then annotated the last seeker utterance in terms of its intent. We categorized these utterances using the following main intents:\footnote{Remember that there are more intents in the dataset, but here we only consider cases where the system responses were scored with ``entirely meaningless'' here.}
\begin{itemize}
  \item \emph{Initiate dialog:} Utterances in this category are for example ones that open the conversation, usually greetings.
  \item \emph{Attribute:}  Here, seekers asked for specific attributes or characteristics of a movie (as discussed also in Section \ref{subsec:failure-specific-question}).
  \item \emph{Recommender's opinion:} This category was used for cases, where the seeker asks the recommender for an opinion, usually regarding a specific movie.
  \item \emph{Neutral (no intent):} These are situations, where no particular recommen\-dation-related intent was found, e.g., when the seeker confirmed the previous statement of the recommender.
\end{itemize}

When annotating these apparently challenging situations, we checked if the very negative score given by the study participant is actually fully plausible or if it might in parts be caused by an overly critical assessment by the participant. We found a few such cases and labeled them as being at least partially ``Subjective''.

In total we found 32 unique situations where responses from the three systems were considered meaningless. The results of our intent categorization are presented in Table \ref{tab:failed-situations}, which shows the number of cases for each intent.  Overall, the results show that both the language generation and our retrieval approach failed in almost 60\% of the situations that end with the user's particular \emph{question}. Similarly, in around 30\% of the situations, these systems seems to have difficulties reacting with a suitable response, where the last seeker utterance does not convey any particular recommendation-related intent, e.g., ``\emph{Yeah, he was Salieri}''.

\begin{table} [h!t]
\renewcommand\thetable{6}
  \centering
\small
\begin{tabular}{p{5cm}ccc}
\hline
 \textbf{Category} & \textbf{Nb. of Cases} & \textbf{Subjective}\\ \hline
 Initiate Dialog & 3 & 3\\
 Attribute & 5 & 0 \\
 Recommendation & 10 & 6 \\
 Recommender's Opinion & 3  & 0 \\
 Neutral & 11 & 3 \\
\hline

\end{tabular}
\caption{Statistics of Challenging Dialog Intents}
\label{tab:failed-situations}
\normalsize
\end{table}

Extending our analysis from Section \ref{subsec:failure-specific-question}, we found that all systems not only often have problems to answer specific questions, but in many cases do not respond appropriately when the seeker has not expressed a particular recommen\-dation-related intent in the last utterance. This in our view points to an interesting research question, i.e., how a system can detect that it has to take the initiative (again) to move the conversation forward, e.g., in case the seeker responds with a confirmation of a previous utterance by the recommender or makes a philosophical statement such as \emph{``well, he's not a real killer if he didn't kill anybody, right?}''.

\subsubsection{Qualitative Analysis of Seeker Preference Expression}
Preference elicitation is a central part of the conversational recommendation process, and various papers in the literature emphasized on how to elicit user preferences, e.g., \cite{ArbpitNavigationByPreference2020, warnestaal2007interview, Sun:2018:CRS:3209978.3210002, DBLP:conf/aaai/YanDCZZL17, Zhang:2018:TCS:3269206.3271776}. In traditional, system-driven approaches, the CRS often takes an active role and asks users about specific preferences, for example, regarding certain item features. However, not so much research seems to exist when it comes to CRS that support free natural language interactions and mixed-initiative dialogs, where the user may also play a more active role. Specifically, in order to build better CRS in the future, it is important to understand in which ways users express their preferences, see also \cite{radlinski2019coached}.

In this work, we therefore conducted a qualitative and exploratory analysis of the ReDial dataset to investigate how users state their preferences to the human recommender. Again, we took the valid cases from our crowdsourced user study and manually inspected the dialogs. We particularly focused on dialog situations that ended with a request by the seeker for a recommendation. In our analysis, we developed a classification scheme consisting of 7 main categories and a number of sub-categories. The classification scheme, along with a number of examples in each category, is shown in Table \ref{tab:preference-category}.

Our preliminary analysis shows that various forms of preference expression can be found. Some seekers mention preferred movie characteristics, others provide the viewing context, some mention reference movies they like, and yet others state what they do \emph{not} like. In some cases, seekers also used a mix of these things, e.g., by providing both a genre and a favorite movie in that genre. Finally, we found situations where the seeker did not specify any particular preferences.

The analysis indicates that correctly understanding these various forms of preference expressions can be very challenging, in particular when there are not too many cases of certain types in the underlying dataset used for language generation or learning. Negations might also be a problem. While the detection of negations in a sentence is a long-studied problem, additional measures may have to be taken in end-to-end learning systems to incorporate such existing techniques. The use of sentiment analysis techniques was explored, for example, in \cite{li2018towards}. Generally, however, more basic research seems to be required beyond the somewhat artificial ReDial dataset to understand the finer nuances of how people communicate in recommendation dialogs, see also \cite{christakopoulou2016towards}.

\begin{table} [h!t]
\renewcommand\thetable{7}
  \centering
\resizebox{!}{145pt}{
\small
\begin{tabular}{p{2.8cm}p{3cm}p{5.5cm}}
\hline
 \textbf{Preference Type} & \textbf{Sub-category} & \textbf{Example} \\
\hline
\multirow{6}{3cm}{Attribute-based}   & Actor-based  &  Hi, can you recommend a good \emph{Resse Witherspoon} movie. Any genre will do. \\
                  & Keyword-based &  Not bad but i want lots of \emph{guts and blood} that was a good one.\\
                  & Genre-based &  Hi, can you recommend a good \emph{romantic comedy}?
                   \\\hline
\multirow{3}{4.2cm}{Context-based} &  &  Hello, I am looking for movies for a night with \emph{friends} that I have coming up.  Any suggestions?
                      \\\hline

\multirow{3}{5.5cm}{Item-based}      &  &  I am looking for a movie a lot like \emph{``Braveheart (1995)"} do you have any suggestions?
                  \\\hline

\multirow{6}{4.2cm}{Negation-based} &  Negating a reference movie &  \emph{``Bride Wars (2009)"} probably isn't my cup of tea Any other suggestions?\\
                &  Negating a keyword &  I'll pass on ``Geostorm (2017)" I hate \emph{disaster} movies Any other ideas?\\

                &  Negating a reference genre &   I like most genres except for \emph{horror} and speculative(\emph{fantasy/science fiction})
                      \\\hline
\multirow{1}{4.2cm}{Novelty-based} &  &  I'm looking for something \emph{new} though
                      \\\hline
& & \\              \hline
\multirow{5}{4.2cm}{Combinations} & Genre-based \& Item-based &   I like \emph{horror} \emph{``The Exorcist  (1973)"} is my favorite. \\
                & Genre-based \& Novelty-based &    I've seen that one too.  It's really funny. Any more \emph{recent comedies} you would suggest?
                      \\\hline
\multirow{2}{4.2cm}{No preference} &  &  Can you give me \emph{any} movie suggestions?
                      \\\hline
\end{tabular}
} 
\caption{Different Ways of Providing Preferences}
\label{tab:preference-category}
\normalsize

\end{table}

\section{Implications \& Discussion}
\label{sec:discussion}
In this section, we summarize the main implications of our research and discuss research limitations.

\subsection{Implications}
Generally, we observed substantial progress in neural language generation approaches to conversational recommendation in recent years. In fact, the analyzed systems (KBRD and KGSF) could react to seeker utterances in a satisfactory manner in many situations. Moreover, as new datasets become available for learning, e.g., \cite{hayati2020inspired,fu2020cookie}, we believe that the performance of learning-based approaches will further increase when the limitations of datasets like ReDial are overcome.

However, our work indicates that retrieval-based approaches, which are barely investigated in the CRS literature, may be promising alternatives to generation-based approaches or could be combined with them in a hybrid system. Such hybrids could help to overcome limitations of the individual approaches and lead to systems that are able to respond to user utterances in even more reliable ways. Besides the use of retrieval-based approaches in hybrid solutions, there is also substantial room for improvement regarding retrieval-based solutions in general. The \crbCRS system proposed in this work relies on traditional similarity measures and ranking heuristics. Many improvements to this first approach are possible, e.g., by using more elaborate ranking techniques or by utilizing more sophisticated techniques for context or intent recognition. Note also the proposed technique can be applied in other domains as well and is not limited to movie recommendations. The application to other domains mainly requires the fine-tuning of parameters and the exchange of application specific background knowledge, like the rating database or the set of keywords (i.e., the genres in our case). The need for such adaptations is however not specific to retrieval-based approaches and adaptation are also required for language generation approaches like KBRD and KGSF, in particular when they rely on domain-specific knowledge graphs.

The additional analyses in our paper furthermore point to the problem that the examined generation-based systems have difficulties to properly react to certain types of user utterances. In particular, they struggle to respond to specific user questions, and they often return inappropriate responses when there is no clear recommendation-related intent in the seeker's utterance. Some of these limitations may have to do with the characteristics of the ReDial dataset, as mentioned above. On the other hand, it still seems important that next-generation CRS have better ways of tracking dialog states and the users' intents. The integration of additional external data sources may be helpful as well when it comes to answering questions about movie metadata. One opportunity for future work in this context may also lie in the integration of general pre-trained language models such as BERT or GPT-3. In \cite{WhatDoesBertKnow2020}, the authors recently analyzed what BERT knows about movies and items in other domains, and concluded that BERT is able to provide content-based information, e.g., about genres, in many cases. In context of returning suitable responses, it is important to note that \emph{coherence} with the ongoing dialog is a highly desirable property of the system's responses. Our proposed approach is able to consider previous utterances in the dialog in the response retrieval phase. And, we also consider coherence as a quality factor in our evaluation when we ask the human judges to score the system responses in terms of their meaningfulness as a continuation of a given dialog situation.

Our work also points to methodological questions related to the evaluation of CRS. Current research on neural CRS models acknowledges that it is important to involve humans in the evaluation process, knowing that pure offline evaluations are usually not sufficient to assess a highly-interactive system like a CRS. However, the descriptions of the evaluations involving human judges in recent papers often appear too superficial, and they sometime do not provide sufficient details about, e.g., design choices for the study, who the judges actually are and what specific task they were given in which environment. We hope that with our study design, we are able to provide a first blueprint regarding how user studies can be made at scale with the help of crowdworkers. Moreover, our additional analyses also focus on the question what does \emph{not} work, and why a CRS may have problems in certain dialog situations. We believe that future evaluations more often should also consider questions when and why systems fail instead of only focusing on improvements in aggregate measures.

\subsection{Research Limitations}
One potential threat to the validity in user studies like ours lies in the reliability of the study participants. Given that we applied more than one quality-assurance measure---participant selection, an attention check, manual inspection---we are confident that our results are reliable. Given also that most participants are fluent in English and regular movie watchers, we believe that the participants are representative at least for a subset of potential users of an online CRS.

Another potential limitation is that we so far only analyzed two language generation systems. The question therefore remains to what extent the findings of our study would generalize to other approaches. Since both analyzed systems (KBRD and KGSF) were published in the last two years, and since they were published at top-ranked scientific conferences, we believe that they are good representatives of the state-of-the-art in neural generation-based systems. Moreover, an earlier analysis of the DeepCRS system in \cite{JannachManzoor2020} indicates that similar phenomena might be found also for other approaches. Note also that our study can be easily extended to include alternative or even newer approaches, as long as the respective authors share the needed artifacts for reproducibility. As of now, we did  not find any work that demonstrate superior quality than the KGSF system, and that the source code of such a system is available. To ensure replicability of our own work, we also share all the code and data used for our analyses online.

So far, we analyzed our approach only with the help of the ReDial dataset. This choice was necessary to ensure a fair comparison with two recent works (KBRD and KGSF), which also relied on this dataset. Recently, a number of alternative dataset was proposed, e.g., \cite{zhou2020towards, hayati2020inspired, liu2020towards}. An evaluation of all compared methods on other datasets is however beyond the scope of our present work, which aimed to assess the relative performance of generation-based and retrieval-based system based on the dataset for which they were originally designed and tuned.

Regarding the general nature of the proposed \crbCRS system, it is a retrieval-based system. However, as discussed in Section \ref{sec:system-details}, we also rely on a small set of heuristics in a few processing phases, which are, for example, implemented using keyword lists and string matching. These heuristics, which are not yet learned automatically, are documented in the provided source code of our system. Automating the construction of these rules and keyword lists, e.g., based on MovieLens metadata, is a part of our future work to avoid any knowledge-engineering bottlenecks.

\section{Conclusion}
Conversational recommender systems (CRS) that interact with users in natural language obtained increased attention in the past few years.  In this paper, we have proposed a retrieval-based approach to conversational recommendations, and we conducted a study involving humans to understand how our system performs compared to recent language generation approaches. Our study led to promising results, and we hope that our study design can be used as a blueprint for user-centric studies of CRS in the future. Overall, the main conclusion of our studies is retrieval-based approaches to CRS can be a promising alternative or complement to language generation approaches.

\bibliographystyle{elsarticle-num}
\bibliography{main}

\end{document}